\documentclass{article}

\usepackage{graphicx}		
\usepackage{textgreek}
\usepackage{pdfpages}
\usepackage{ccaption}

\begin{document}

\centerline{A SELEX Protocol to Identify ssDNA Biotemplates for Gold Nanoparticle Synthesis}

\vspace{1cm}

\author{Philip Calabrese, PhD}

\begin{abstract}

A modified ssDNA SELEX protocol was developed in order to evolve a randomized library of imidazole modified ssDNA sequences towards sequences that mediate the formation of gold nanoparticles from Au\textsuperscript{3+} precursor ions in aqueous solution.  Active sequences bound to nanoparticles were partitioned from inactive sequences based on density via ultracentrifugation through a discontinuous sucrose gradient.  Colloidal gold solutions produced by the evolved pool had a distinct absorbance spectra and produced nanoparticles with a narrower distribution of sizes compared to colloidal gold solutions produced by the starting, randomized pool of imidazole modified ssDNA.  Sequencing data from the evolved pool shows that conserved 5 and 6 nt motifs were shared amongst many of the isolates, which indicates that these motifs could serve as chelation sites for gold atoms or help stabilize colloidal gold solutions in a base specific manner.  

\end{abstract}

\vspace{.7cm}

INTRODUCTION

\vspace{.7cm}

Biomolecules that assemble nanoparticles from precursor ions offer new ways to address limitations in the bottom up synthesis of nanoscale materials as well as improvements upon methods used in the investigation of biological systems.  For example, the ferritin protein, whose function it is to store iron within a core of 24 protein subunits, has been used as an electron dense tag to follow proteins in electron microscopy. ~\cite{wang2011ferritin}.    Biomolecules that can template nanoscale crystallization processes at physiological pH and temperature have also been used as parts in bottom-up assembly strategies for hybrid nanoscale materials that can be programmed to position biological components into hierarchical assemblies.  For example, the Belcher lab at MIT has created new routes for the assembly and synthesis of nanowires for lithium ion battery electrodes using genetically engineered viruses with high affinity sites for certain components of the battery that template nanoparticles.  By incorporating gold-binding peptides into the filament coat, they were able to form hybrid gold–cobalt oxide wires that improved battery capacity ~\cite{lee2009fabricating}. 

A variety of examples which utilize nucleotides and nucleic acids as templates for the synthesis of a variety metal nanoparticles exist.  These include tRNA templated CdS quantum dots ~\cite{ma2006rna}, nucleotide capped gold nanoparticles ~\cite{zhao2007highly}, and Ag nanocluster formation templated by cytosine oligonucleotides ~\cite{ritchie2007ag}.  All of these approaches require high concentrations of metal precursor and template nanoparticles in non-sequence specific ways.     With the impressive amount of progress that has been made regarding gold nanoparticle synthesis and functionalization, few examples exist which utilize evolutionary methods to discover sequence specific ssDNA biotemplates for gold nanoparticle assembly.  

One route towards the discovery of sequence specific oligonucleotide templates that assemble metal or metal oxide nanoparticles is the use of Systematic Evolution of Ligands by Exponential Enrichment (SELEX).  In this approach, nucleic acid sequences which remain bound to nanoparticles after an incubation step are partitioned from unbound sequences and amplified with PCR (Figure 1).

\begin{figure} [h!TB]
    \includegraphics[width=0.9\textwidth] {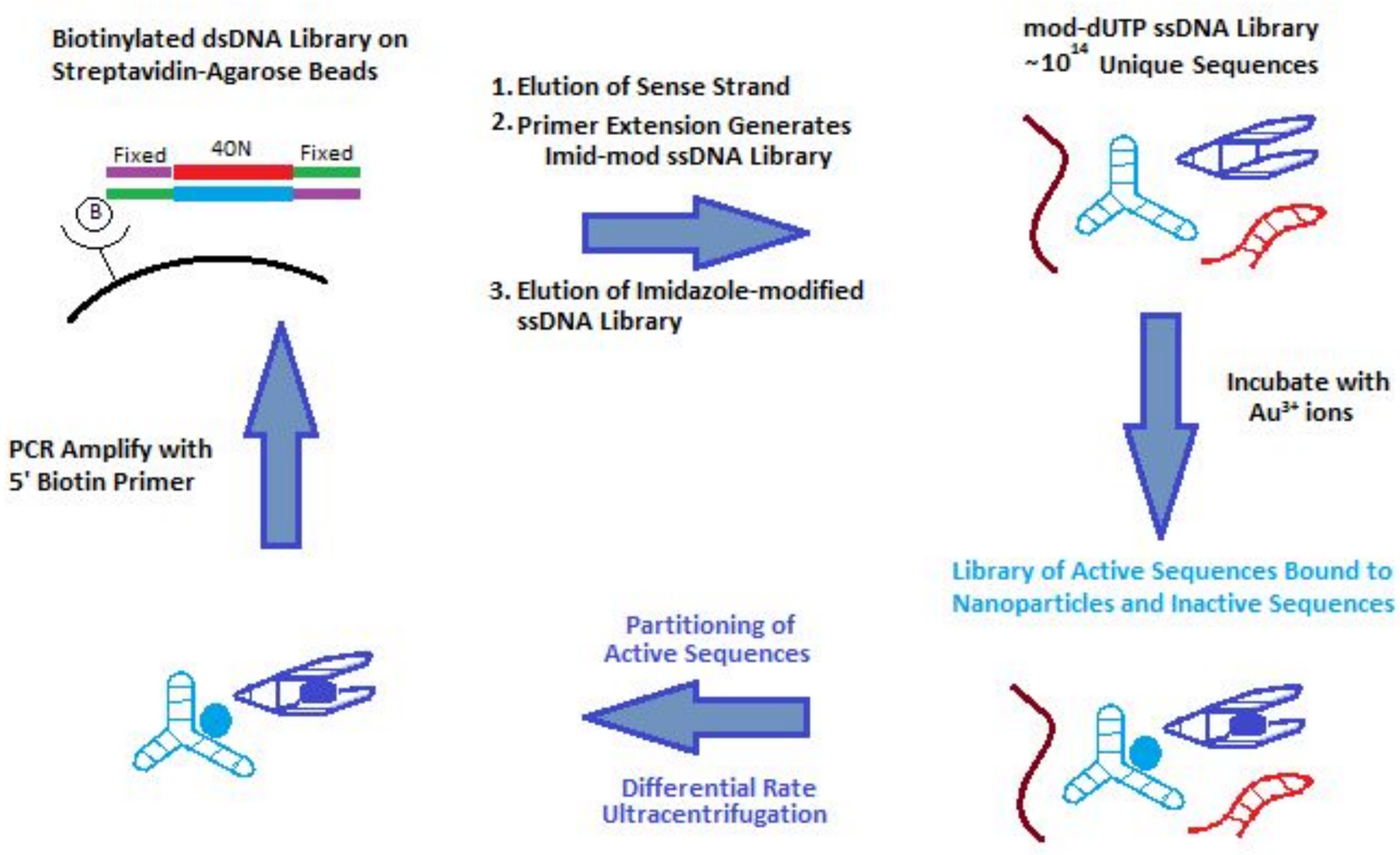}
    \label{In-vitro SELEX Scheme for Isolating Nanoparticle Templating Sequences}
     \caption{Modified ssDNA SELEX scheme for nanoparticle templating sequences.}
\end{figure}

 The captured sequences are exponentially enriched and regenerated into the next round's starting population where they are taken through another iteration of selection and amplification.   In the first SELEX based experiment of its kind, RNA sequences capable of mediating the formation of palladium and platinum nanoparticles from metal precursor ions found through \emph{in-vitro} SELEX were found to assemble nanoparticles with control over shape and crystallinity ~\cite{gugliotti2004rna}.  An advantage of using SELEX for nanoparticle synthesis is that evolved sequences can work with low concentrations of metal precursor, which at higher concentrations might otherwise precipitate with other components of a biological sample.

Previous reports have implicated histidine as a crucial residue in the templating of gold nanoparticles with certain peptide sequences that were used in bionanomaterials synthesis applications ~\cite{djalali2002nanowire}.  For example, the  repeating consensus sequence AHHAHHAAD from the histidine-rich protein II of \emph{Plasmodium falciparum} was reported to mediate the aqueous self-assembly of zerovalent gold metal clusters with the aid of a reducing agent.  This same sequence was subesquently shown to mediate the assembly of gold nanowires after being immobilized on heptane dicarboxylate nanowires.  The imidazole side chain offers an electron donating nitrogen that has been shown to form coordination complexes with Au\textsuperscript{3+} and other metal cations ~\cite{de2007synthesis}.  Due to imidazole’s tendency to form coordination complexes with Au\textsuperscript{3+} and perhaps play a direct role in peptide mediated gold nanoparticle synthesis, imidazole modified dUTP (Figure 2) was incorporated at every uracil base in the random region of the starting library used in the modified ssDNA SELEX sheme.

\begin{figure}[h!tb]
\center
   \includegraphics[width=0.5\textwidth]{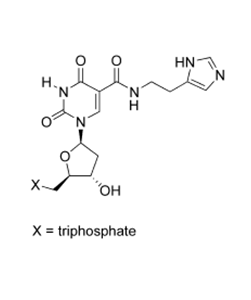}
    \caption{Structure of imidazole modified dUTP.}
    \label{}
\end{figure}

In a model proposed by Berti, et. al, nucleation processes that are triggered by nucleotides or nucleic acids begin when a precursor complex forms between a metal ion and the nucleic acid.  At physiological pH, positively charged metal cations interact with negatively charged phosphate groups through simple, non-specific electrostatic attraction. Alternatively, empty orbitals of metal cations can accept electrons from basic amino and keto groups within nucleobase moieties.  This leads to the coordination of metal complexes which, if positioned correctly in relation to one another in the zerovalent state, could lead to the formation of metal metal-bonds and the establishment of critical nuclei.  After a critical nuclei of atoms are established, a growth phase ensues whereby individual atoms begin to deposit on the nanoparticle surface.   Eventually, stabliziing interactions between the biomolecule and capping ligand become more thermodynamically favorable than the addition of new atoms to the nanoparticle surface and if the conditions are right, the nanoparticles in solution will remain stable as the biotemplate remains bound to the surface ~\cite{Berti2008Nucleic}.

The first materials based RNA \emph{in-vitro} selections reported upon in the literature partitioned sequences bound to nanoparticles from unbound sequences by centrifuging RNA/metal precursor incubations through100K MWCO regenerated cellulose membranes in order to flow through unbound sequences and retain sequences bound to nanoparticles ~\cite{gugliotti2004rna}.    These membranes are not ideal for SELEX partitioning steps since nucleic acids can non-specifically stick to the membrane and ultimately confound the results of the selection.  Approaches for isolating RNA sequences bound to iron oxide nanoparticles based on the material's susceptibility to a magnetic field were also reported where the authors selected for sequences based on a desired property of the material ~\cite{carter2009vitro}.  Selection based on a material's property is an attractive idea but not all types of nanoparticles lend themselves to this approach. 

One common theme among metal nanoparticle - nucleic acid templated processes is that the selected nucleic acids stay bound to the nanoparticle surface.  We set out to develop a rapid and robust protocol for the isolation of ssDNA sequences that template nanoparticles using ultracentrifugation through a discontinuous sucrose gradient as a means of separation.  Using this method, we were able to effectively partition active sequences from inactive sequences based on the difference in sedimentation velocity experienced by the DNA sequences bound to highly dense gold nanoparticles compared to unbound, inactive DNA sequences.

\vspace{.7cm}

EXPERIMENTAL AND RESULTS

\vspace{.7cm}

\noindent
\textbf{
\emph{Development of SELEX Partitioning Step}}

\vspace{.7cm}

The parameters of the SELEX experiment were set so that selected sequences could work at physiological pH using sub micromolar concentrations of Au\textsuperscript{3+} precursor.   A mild reducing agent, L-tyrosine, was included at 1 mM to facilitate reduction of Au\textsuperscript{3+} to Au\textsuperscript{0}.  Control experiments indicated that L-tyrosine alone at 1 mM concentrations can not stabilize colloidal gold solutions over prolonged periods of time.

In order to partition active sequences from inactive sequences, we used differential rate ultracentrifugation through a discontinuous sucrose gradient.  This method of separation is commonly used in ultracentrifugation of high density particles such as metal nanoparticles and viruses.  Since density determines sedimentation rate, we saw the opportunity to separate gold nanoparticles from 100 nt single stranded nucleic acid sequences.  Since sedimentation velocity is dependent on the density of a molecule in solution, a gradient can be designed such that the bottom layer provides a "cushion" whereby the high density particles in solution sediment into a narrow band at the interface of the discontinuous gradient. To achieve separation in general, the high density particles, which in our experiment are the ssDNA capped gold nanoparticles, should move quickly through the top layer of the discontinuous gradient and halt at the more dense layer of sucrose.  Meanwhile, the low density particles, unbound inactive ssDNA, do not migrate into the top layer of sucrose.

In order to determine the sucrose layer composition as well as the optimal times, ultracentrifugation speeds and temperatures to use in the SELEX partitioning step, 5 nm diameter gold nanoparticles protected with a self assembled monolayer of triethylene glycol mercaptoundecanoic ether were used as a standard.  The deep red colored colloidal gold could be visually monitored before and after the centrifugation step in order to confirm their migration to the sucrose cushion.  A 30\% over 67\% discontinuous sucrose gradient was poured in a Beckman SW-40 polyallomer centrifuge tube with 3.75 mL 30\% sucrose layered on top of 1 mL 67\% sucrose.   A 200 uL sample of 10 uM of PEG$_3$ alkane thiol protected 5 nm diameter gold nanoparticles was applied to the top of the gradient.  The tubes were loaded into a Beckman SW 40-Ti swinging bucket rotor and spun in an ultracentrifuge at 30,000 rpm for 30 minutes, 25 C.   After the ultracentrifugation step, PEG$_3$ alkane thiol protected 5 nm diameter gold nanoparticles sediment into a visible, relatively narrow band at the boundary of the gradient (Figure 3).

\begin{figure}[h!tb]
  \begin{center}
    \includegraphics[width=0.6\textwidth]{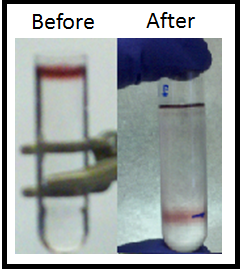}
\end{center}
    \caption{TEG-MUE protected gold nanoparticles sediment to the boundary of a 30\% over 67\% sucrose gradient.}
    \label{}
\end{figure}

We found that the most effective method for sample recovery was puncturing the side of the tube with an 18G syringe needle attached to a 3 mL non-stick plastic syringe and extracting out the material above from just slightly below the interface of the original gradient where the nanoparticles had sedimented.   This method provided the most robust and reproducible way to extract sequences bound to gold nanoparticles from round to round during the selection.

Before the ultracentrifugation step, a mark was placed on the side of the tube 2 mm below the interface of the discontinuous gradient to indicate exactly where to puncture the tube at the end of the separation.  200 uL of 10 uM PEG$_3$ alkane thiol protected gold nanoparticles was layered on top of the gradient, carefully placed in a swinging bucket rotor and ultracentrifuged at 30,000 rpm for 30 minutes.  After the spin, 500 uL of sample was extracted from above the interface of the gradient and buffer exchanged into 1X PBS using 100K MWCO membrane spin filters.  The yield was quantified using a UV-vis spectrophotometer.  Using the peak observed at 535 nm, we calculated a 56\% recovery of TEG-MUE protected gold nanoparticles.  The low yield could be attributed to nanoparticles that were lost to the side of the pollyallomer ultracentrifuge tube, the pores of the 100K MWCO membrane and to the syringe.

The migration of a ssDNA 84-mer was tested through a 30\% over 67\% discontinuous sucrose gradient in order to ensure that unbound oligonucleotides would remain on top of the 30\% sucrose and not amplify from the region where gold nanoparticles would be extracted from. 
\begin{figure}[h!tb]
  \begin{center}
    \includegraphics[width=0.9\textwidth]{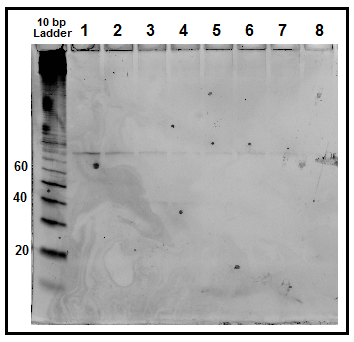}
\end{center}
    \caption{8\% denaturing PAGE gel of 500 uL fractions from the top of the gradient.}
    \label{}
\end{figure}

 A 200 uL sample of 5 uM ssDNA (one nanomole) was loaded on top of the sucrose gradient and spun at 30K RPM for 30 minutes.  Successive 250 uL fractions were removed from the top of the gradient, buffer exchanged into 1X PBS buffer using a 10K MWCO membrane and concentrated to 10 uL total volume using a speed-vac apparatus.  Each successive fraction was placed on an 8\% denaturing PAGE gel with the fraction closest to the top of the original gradient corresponding to lane 1 (Figure 4).   The detection limit of the gel stain used in this experiment was about 10 - 100 ng of ssDNA.

\vspace{.7cm}

\noindent
\textbf{
\emph{Determining Reducing Agent Strength and Buffer Composition}}

\vspace{.7cm}

An appropriate buffer and reducing agent were chosen to use during incubation of an imidazole modified ssDNA 40N library with HAuCl$_4$ in aqueous reaction conditions.  Buffer choice is an important consideration not only because of the pH and ionic strength of the reaction environment but also because the components of the buffer must not facilitate nucleation and growth of nanoparticles on their own.  Because of this, common buffers that contain compounds with amino groups and heterocylcles were avoided due to the direct role these groups could potentially contribute towards nanoparticle nucleation and growth processes.  We decided upon phosphate buffered saline (PBS) without magnesium or calcium because this buffer is generally amenable to biological protocols and should remain relatively inert towards the reaction.  We did not include calcium or magnesium because this would also eliminate the chance of evolving sequences that incorporate these metals into nanoaprticles.

Some amino acids and common biological reducing agents have been reported to serve as stablizing and capping agents in colloidal gold synthesis protocols.  We intended upon choosing one at a concentration that would facilitate the reduction of Au\textsuperscript{3+} to Au\textsuperscript{0} but not compete with ssDNA as a capping agent.  We chose a variety of biological reducing agents that had measured redox potentials which would indicate that it was thermodynamically favorable to reduce Au\textsuperscript{3+} to Au\textsuperscript{0} at physiological pH.  We experimented with NADH, citrate, oxaloacetate, ascorbic acid and L-tyrosine at 1 mM and 10 mM concentrations in order to determine which could facilitate the reduction of Au\textsuperscript{3+} to its zerovalent state but not produce stabile solutions of colloidal gold on their own.  To do this, we monitored the solutions absorbance in the 500 - 600 nm range to see if the reducing agent alone could stabilize colloidal gold solutions in 1X PBS supplemented with 400 uM HAuCl4.   

\begin{figure}[h!tb]
  \center
    \includegraphics[width=0.9\textwidth]{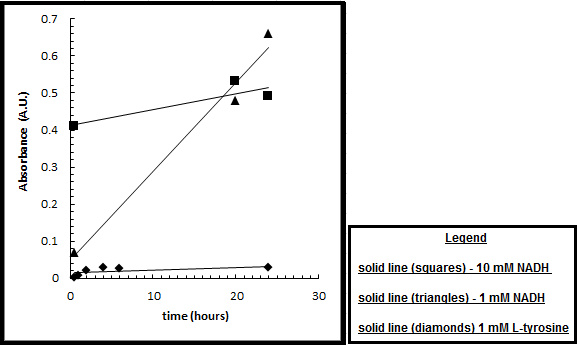}
    \caption{Common biological reducing agents facilitate the reduction of Au\textsuperscript{3+} to Au\textsuperscript{0}.}
    \label{}
\end{figure}

The optical density at 550 nm is plotted over time in the various HAuCl4/reducing agent mixtures (Figure 5).  10 mM and 1 mM NADH quickly reduced Au\textsuperscript{3+} to Au\textsuperscript{0} as evidence of colloidal gold emerged within the first 30 minutes of incubation time. Ascorbic acid, citrate and oxaloacetate produced similar results and did not lend themselves suitable for our application. We did not want to use reducing agents that would facilitate the assembly of gold nanoparticles within hours of incubation time.   Ultimately, we chose L-yrosine at a 1 mM concentration because when combined with 1X PBS in 400 uM HAuCl4, we noticed the slow reduction of Au3\textsuperscript{3+} to Au\textsuperscript{0} that did not result in a solution with strong absorbance peak anywhere in the 500 - 600 nm range as evidenced by a mild absorbance peak at 550 nm which plateaued over time.

\vspace{.7cm}

\noindent
\textbf{
\emph{Imidazole Modified ssDNA Library Generation}}

\vspace{.7cm}

In order to generate a ssDNA library with a 40N randomized region containing imidazole-modified dUTP at every uracil position in the random region, we used previously described methods that utilize Deep Vent (exo-) polymerase to incorporate modified dUTP during primer extension of a randomized antisense ssDNA template library immobilized on streptavidin agarose ~\cite{vaught2010expanding}.  Deep Vent (exo-) polymerase is able to incorporate this modified dUTP with high efficiency and generate full length product at greater than 95\% yield.  

\vspace{.7cm}

\noindent
\textbf{
\emph{Incubation of Imidazole Modified ssDNA Library with HAUCl4}}

\vspace{.7cm}

After generating one nanomole of imidazole modified ssDNA library for the SELEX experiment, 350 picomoles of ssDNA library containing a 40 nt random region was dissolved in 0.1 M Phosphate Buffer, pH 7.4 with 1 mM L-Tyrosine to a final volume of 1 mL in a DNAse/RNAse free, non-stick eppendorf tube.  An identical reaction without L-tyrosine was also prepared in order to see if imidazole modified ssDNA could facilitate the production of gold nanoparticles without a reducing agent.  The ssDNA containing solution was heated to 75 °C for 2 minutes to induce thermal melting of the ssDNA molecules and the the reaction mixtures were allowed to slowly cool to ambient temperature to induce re-folding.  0.5 uL of 0.8 M aqueous gold (III) chloride was added to a final concentration of 400 uM.   As a negative control, a reaction mixture that contained no ssDNA but all other components was prepared.  The reaction mixtures were vortexed briefly and incubated at room temperature for 17 hours.  Figure 6 shows that the reaction containing imidazole modified ssDNA library (blue) exhibits a strong absorbance peak at 535 nm, indicating the presence of gold nanoparticles. In comparison, the reaction without ssDNA (green dash) absorbs weakly throught the 500 - 600 nm range without a comparitively defined peak.

\begin{figure}[h!tb]  
\center
    \includegraphics[width=0.6\textwidth]{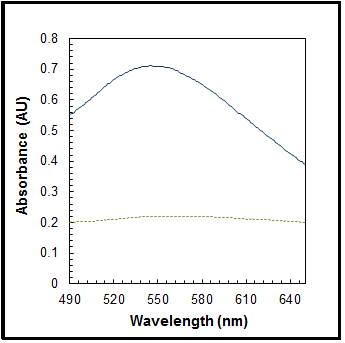}
    \caption{Round 0 imidazole modified DNA library(blue) vs. buffer(green).}
    \label{}
\end{figure}

\vspace{.7cm}

\noindent
\textbf{
\emph{SELEX Partitioning Step}}

\vspace{.7cm}

After 17 hours of incubation, the reaction mixture containing gold nanoparticles and imidazole modified ssDNA library was layered on a 30\% over 67\% discontinuous sucrose gradient and spun at 30K rpm for 30 minutes, 25 C.   A sample containing the same volume and concentration of imidazole modified ssDNA library without Au\textsuperscript{3+} ions was run opposite of the reaction incubation to ensure that unbound ssDNA sequences would not PCR amplify from the interface of the gradient under identical ultracentrifugal forces.

Gold nanoparticles that sedimented to the boundry of the discontinuous gradient were collected by puncturing the side of the tube with an 18G syringe needle. 500 uL of sample was drawn out from 2 mm below the original interface of the gradient.  The extracted sample was concentrated on a 100K MWCO membrane and buffer exchanged into 1X PBS to a final volume of 75 uL.   In order to amplify sequences bound to gold nanoparticles, PCR reactions containing a 3' biotinylated antisense primer were prepared in triplicate with 15 uL of the final 75 uL sample volume derived from the boundary of the sucrose gradient.  50 uL PCR reactions containing NEB HotStart Taq polymerase and all other PCR reaction components was prepared.  A 5 uL aliquot of the combined PCR reactions was run on an 8\% denaturing PAGE gel to ensure that the PCR product was the proper length and not a result of primer dimer.  The incorporation of biotin to the antisense strand was also confirmed by using streptavidin to shift the band into a higher molecular weight complex.

The PCR amplified, biotinylated dsDNA was captured on 30 uL of streptavidin-agarose resin (50/50 slurry) and the native sense strand was eluted from the beads using 20 mM NaOH. The streptavidin-agarose beads were washed five times with 1X PBS buffer before being resuspended in 1X SQ10 buffer.  Imidazole modified ssDNA was then enzymatically synthesized by performing a primer extension reaction on the captured biotinylated antisense strand using Deep Vent (exo-) polymerase.  The modified ssDNA was eluted from streptavidin agarose with 20 mM NaOH and the pH of the eluant containing ssDNA was neutralized with 1/4 volume 80 mM HCl.  Before entering the next round of selection, the length of the modified ssDNA was determined using an 8\% denaturing PAGE gel in order to ensusre the synthesized strand was of proper length.  This selection cycle was repeated for 8 rounds while sequentially lowering incubation times and concentrations of gold precuror ions (Figure 7)

\begin{figure}[h!tb]
  \begin{center}
    \includegraphics[width=0.7\textwidth]{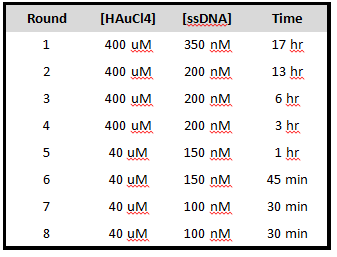}
\end{center}
    \caption{Conditions of SELEX experiment in each round.}
    \label{}
\end{figure}

Figure 8 (left panel) shows the PCR amplification results from round 1.  The group of three traces with an inflection point at cycle 16 were identical aliquots of a reaction incubation with imidazole modified ssDNA library, 400 uM HAuCl4 and 1 mM L-Tyrosine.  The group of three traces that did not cross the threshold was from a sample containing imidazole modified ssDNA library and 1 mM L-tyrosine but not HAuCl$_4$.  Lane 1\% of the 8 denaturing gel is a 10 bp ladder with the lowest band corresponding to 20 bp, the two bands in land 2 contains the anti-sense biotinylated strand of the dsDNA which runs higher than the sense, non-biotinylated strand in an 8\% denaturing PAGE gel.  Addition of streptavidin to the sample in lane 2 shifts only the biotinylated strand into a higher MW complex with one, two, three or four streptavidin subunits attached.  The resultant biotinylated dsDNA from the PCR reactions containing samples from the HAuCl4, 1 mM L-tyrosine reactions were pooled, captured on streptavidin Ultra-Link Resin (Pierce) and the non-biotinylated, native sense strand was eluted from the beads using 20 mM NaOH.  A primer extension reaction was prepared to generate round 2 imidazole modified ssDNA library.

\begin{figure}[h!tb]
  \begin{center}
    \includegraphics[width=0.9\textwidth]{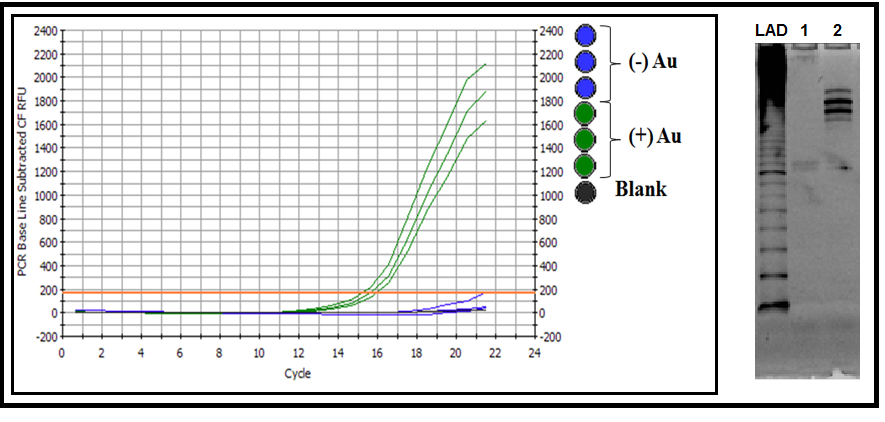}
\end{center}
    \caption{PCR trace of +/- Au Samples (left), 8\% denaturing PAGE gel  of PCR reaction (right).}
    \label{}
\end{figure}

Reactions containing 350 nM Rd. 0 imidazole modified ssDNA, 1 mM L-tyrosine and 400 uM HAuCl4 turned visibly red within 6 hours of incubation time.  A broad absorbance trace in the 500-650 nm range with a peak at 540 nm was observed in Figure 6, indicating the presence of nanometer sized gold particles. Reaction mixtures without ssDNA began to turn purple with a black precipitate, indicating the slow reduction of Au\textsuperscript{3+} to macroscopic gold precipitate.  Reactions with imidazole modified ssDNA  ultimately produced solutions of a deep red intensity and were stable over time with no precipitate seen after weeks of incubation at room temperature.   Samples without ssDNA turned colorless and had a visible black precipatate after 48 hours, indicating the presence of bulk gold.  

\begin{figure}[h!tb]
  \begin{center}
    \includegraphics[width=0.6\textwidth]{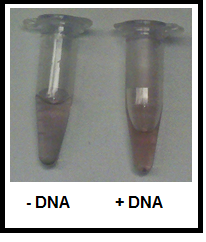}
\end{center}
    \caption{Samples with and without ssDNA after 6 hours.}
    \label{}
\end{figure}

After eight rounds of selection and amplification, the round 8 evolved pool was sent for sequencing and the data were analyzed with Daughter of Sequence Alignment (DOSA) software (SomaLogic).  A variety of conserved 5 nt and 6 nt motifs are found throughout 7 families within which many of those same conserved motifs present themselves (Figure 10).  This could mean that some of these 5 - 6 nt motifs have an affinity for gold ions or for certain surface facets of a growing gold nanoparticle surface.  The conserved n-mers are arbitrarily color coded in the figure.  

\begin{figure}[h!tb]
  \begin{center}
    \includegraphics[width=0.9\textwidth]{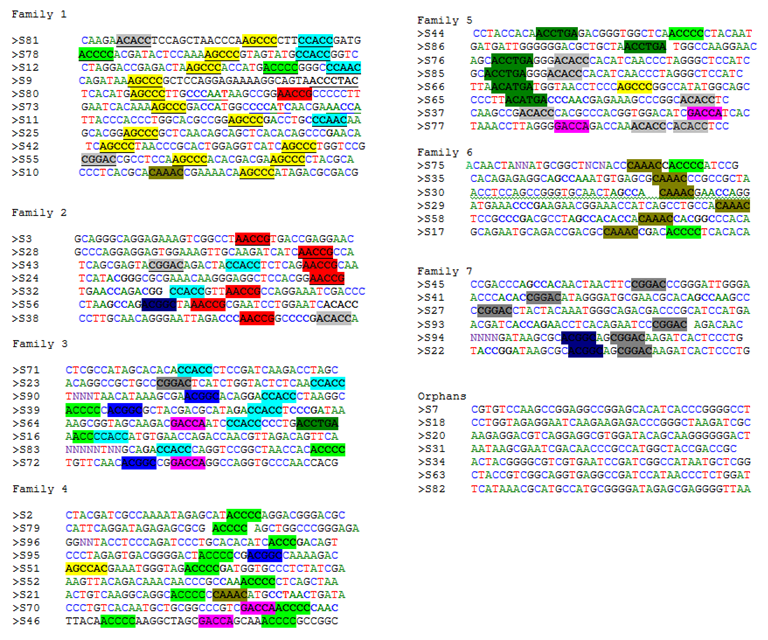}
\end{center}
    \caption{Consensus sequences grouped into families amongst evolved round 8 pool.}
    \label{}
\end{figure}

5 uL of sample from the reaction mixtures used in round 0 and round 8 of the selection were prepared for TEM analysis along with a no DNA control. To prepare samples for TEM, 5 uL of the nanoparticle containing solution was drop cast onto a carbon coated copper (400 mesh) TEM grid (Ted Pella) and allowed to dry under ambient conditions. (Figure 11)

\begin{figure}[h!tb]
  \begin{center}
    \includegraphics[width=0.9\textwidth]{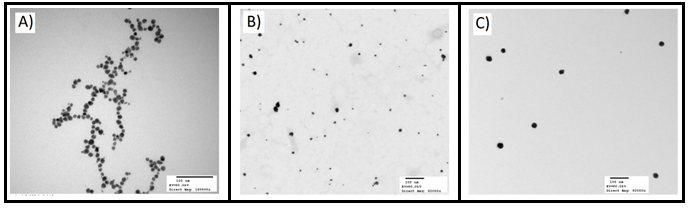}
\end{center}
    \caption{TEM images of round 8 imidazole modified ssDNA (A), round 0 imidizole modified ssDNA library (B), and no ssDNA (C).}
    \label{}
\end{figure}

Characterization of the morphology of the ssDNA-AuNP conjugates was performed on a Phillips CM10 bright field TEM. The diameter of gold nanoparticles that were imaged by TEM were measured using the ImageJ software measurement tool. Grids prepared with gold nanoparticles templated by round 8 and round 0 imidazole modifed DNA were covered in material. Gold noparticles were less abundant in the no DNA control and were larger in size by comparison with edges that were less defined.

\begin{figure}[h!tb]
  \begin{center}
    \includegraphics[width=0.9\textwidth]{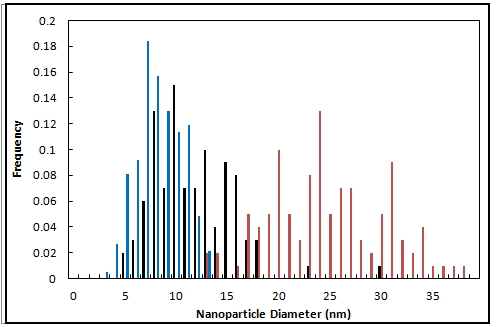}
\end{center}
    \caption{Size distribution of gold nanoparticles amongst rd.8 (red), rd. 0 (blue) and no DNA (black).}
    \label{}
\end{figure}

\begin{figure}[h!tb]
  \begin{center}
    \includegraphics[width=0.9\textwidth]{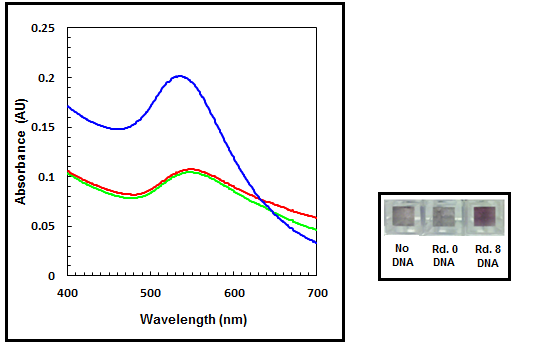}
\end{center}
    \caption{Absorbance spectra comparing no DNA (red), 100 nM rd. 0 DNA (green), 100 nM rd.8 DNA (blue). A picture of the cuvettes used in the measurement are shown on the right.}
    \label{}
\end{figure}

\vspace{5 cm}

CONCLUSIONS AND FUTURE DIRECTIONS

\vspace{.7cm}

We found that imidazole modified ssDNA sequences are capable of templating goldnanoparticles from HAuCl4 and that conserved sequence motifs can be isolated through iterative cycles of selection and amplification. UV-vis spectroscopy of Rd. 0, Rd. 8 and no DNAcontrol indicates that the evolved pool produces gold nanoparticles in greater abundance that are more narrowly dispersed in solution, smaller in diameter compared to an HAuCl4 solution treated with the reducing agent alone or the Rd. 0 starting material. It should also be noted that in this experiment, the reaction volume was one-fifth and the concentration of ssDNA was nearly four fold lower than that of the Rd. 0 incubation. TEM analysis concurs with UV-vis spectroscopy revealing that the evolved pool produces gold nanoparticles with an average diameter of 8.5 nm (+/-) 2.2 nm compared to the Rd. 0 library which produced gold nanoparticles with an average diameter of 12.0 nm (+/-) 4.2 nm.

Ultracentrifugation as a method of separation should be further explored as a partitioning step in similar nucleic acid based materials selection experiments. Our results indicate that ssDNA can serve as both a general template and a sequence specific templatate for the assembly of gold nanoparticles from gold precursor ions. We show that SELEX can be used to find DNA sequnce motifs that assemble nanoparticles from Au3+ which was kept at significantly lower concentrations compared to methods which employ non-specific sequences. 

Sequences that can work with lower concentrations of precursor are advantageous for several reasons. When considering their use in biological applications,the nonspecific precipitation of gold nanoparticles should be less prevalent in assays containing complex mixtures of biomolecules. Towards energy or materials applications, the use of less metal precuror offers huge savings in potential scale-up costs of these processes. 

The evolved motifs found in this selection could potentially be used as an identification tag in electron
microscopy or an optical readout in a biosensor based assay if used as a fusion sequence to a protein binding aptamer whereby the formation of gold nanoparticles from Au3+ ions is exclusively templated by the evolved sequence. One could also imagine using these sequences as a fusion to specifically place gold nanoparticles on nanoengineered structures based on established techniques within the field such as DNA hybridization.

However, further optimization of the selected sequences needs to be completed to achieve these goals. When isolate sequences were tested, we saw evidence that concurred with previous characterizations of RNA materials SELEX outcomes which indicated that sequences work together to effect the outcome of particle crystallinity and morphology. Based on these observations, it seems as if a co-evolution of sequences will occur during materials based SELEX experiments. 

\vspace{1cm}

MATERIALS AND METHODS

\vspace{.7cm}

\noindent
\textbf{
\emph{ssDNA library Design, Synthesis and Purification}}

\vspace{.7cm}

The ssDNA library used in the gold nanoparticle selection scheme was constructed as follows:  

79-mer ssDNA library 5'-ATA TAT ATA CCG AGC ACT GAG TTT GCC - 40N - GCG AAA CGA CAA GAA GAC AAA AAA AA-3', 
5' Forward Primer - ATA TAT ACC GAG CAC TGA GTT TGC C, 
3' Reverse Biotinylated Primer - Biotin - TTT TTT TTG TCT TCT TGT CGT TTC GC. 

The 40N random ssDNA library flanked by constant priming regions was synthesized on the lab's ABI-394 synnthesizer.  DNA phosphoramidites were diluted in dry acetonitrile in an argon purged glovebox and placed on the lab ABI-394 DNA synthesizer at a final concentration of .1 M.  A separate random bottle mixture of dA, dG, dC and dT phosphoramidites was prepared at respective concentrations such that each base had a 1:1:1:1 coupling efficiency during coupling steps of the 40N random region.  In 15 mL of acetonitrile, .363 g of dA, .227 g of dT, .259 g of dG and .387 g of dC were dissolved to make the random mixture bottle   5 g of fresh dicyano imidazole (DCI) was diluted to a final concentration of .25 M in anhydrous acetonitrile and placed on an orbital shaker with molecular sieves for 48 hours prior to DNA synthesis.  7\% dichloroacetic acid was prepared in dichloromethane and placed on the DNA synthesizer with fresh bottles of oxidizing solution, Cap A and Cap B mix (Glen Research).  The resulting ssDNA library was cleaved from the silica support using concentrated ammonia which was subsequently evaporated off using a cold trap vacuum.  The resultant white precipitate was dissolved in Milli-Q H20, pH 7.0.  Full length oligonucleotides were gel purified using 8\% denaturing PAGE.  The band corresponding to the full length synthesis product was crushed into a fine slurry with a pipette tipe in a non-stick eppendorf tube and passively eluted in Milli-Q H$_2$O pH 7.0.   In order to remove excess salts, the ssDNA solution was flowed over G25-Sephadex resin and fractionated into 100 uL aliquots.  An 8\% denaturing PAGE gel containing each fraction was run in order to identify which fractions to pool that contained ssDNA.  The concentration of the pooled fractions was determined by the A$_260$ reading of a 1:50 dilution of the stock sample.  5' and 3' Primers were obtained from IDT DNA and gel purified using an 8 denaturing PAGE gel purification. The excized bands were crushed into a slurry and passively eluted into .3 M NaOAC, 2 mM EDTA overnight. 

\vspace{.7cm}
\noindent
\textbf{
\emph{Biotinylated dsDNA library generation}}

Gel purified ssDNA library containing a 40N random region was subjected to 2-cycle PCR using a 3’ biotinylated primer.  The initial melting step was set to 94 C for 2 minutes, the annealing step was set to 55 C for 30 seconds and the extension step was carried out at 71 C for 90 seconds.  Each subsequent melting step was carried out at 94C for 30 seconds.  The resulting biotinylated dsDNA library was captured on Streptavidin-Agarose beads.  The complementary strand was then eluted off of the dsDNA library by adding 200 uL of 20 mM NaOH.  The beads were pelleted, the supernatant was removed and beads were washed five times with 1X PBS.

\vspace{.7cm}
\noindent
\textbf{
\emph{Primer Extension Reaction to Generate Imidazole Modified ssDNA Library}}

All of the reaction components of a 1X Primer Extension Reaction (NEB) were added to 15 uL streptavidin agarose beads coupled to the ssDNA library template strand, heated to 95 C for 30 s, annealed at 55 C for 5 minutes and incubated at 70 C with orbital shaking at 1000 rpm for 30 minutes.  The supernatant was collected and the beads were washed 3X in 1X PBST and the newly synthesized strand containing imidazole modified ssDNA library was eluted from its complementary strand with 20 mM NaOH.  The beads were vortexed briefly and spun at 4,500 rpm for 15 seconds.  The supernatant containing modified ssDNA library was removed and  ¼ volume of 80 mM HCl was added to neutralize the pH.  The beads were washed 3X in 1X PBST and stored at 4 C with .05 \% Sodium Azide.    ssDNA was quantified by measuring the absorbance at 260 nm on a spectrohotometer before evaporating the solution to dryness.

\vspace{.7cm}
\noindent
\textbf{
\emph{Preparation of PEG$_3$ alkane thiol protected gold nanoparticles}}

Citrate stabilized gold nanoparticles with a covariation variance of <15 and a mean diameter of 5 nm were treated with 1 mM triethylene glycol mono-11-mercaptoundecyl ether (TEG-MUE) in a 50 ethanol solution in order to create PEG3 alkane thiol stabilizing monolayer around the nanoparticle surface.  Gold nanoparticles were placed on an orbital rocker overnight in the 1 mM TEG-MUE ethanol solution.  Excess TEG-MUE and ethanol were removed by centrifuging the solution at 13K RPM in order to sediment the nanoparticles at into a visible, deep red concentrated pellet.  The supernatant was removed and the nanoparticles were washed 5X with pH 7.0 Milli-Q H20 until there was less than 1 ethanol in solution by volume.  The concentration of TEG-MUE protected gold nanoparticles was determined using their absorbance maximum at 520 nm and their cited extinction coefficient at 520 nm.  A critical coagulation test was used to visually confirm the stability of the PEG3 alkane thiol protected nanoparticles versus citrate stabilized particles without a capping agent in 1X PBS.

\vspace{.7cm}
\noindent
\textbf{
\emph{Sequencing of Rd. 8 Evolved Pool}}

TOPO TA vector (Invitrogen) was used to transfect One Shot TOP10 competent E. coli cells to prepare single colonies that were selected based on ampicillin resistance which is conferred to the cells with the insertion of PCR product in the TOPO TA vector.   The general protocol from the manufacturer was followed.  2 uL of fresh PCR product from the Rd. 8 selection step was combined with 1 uL salt solution (1.2 M NaCl, 60 mM MgCl2), 1 uL TOPO TA vector, and 2 uL of Milli-Q H20 to a final volume of 6 uL. 2 uL of the TOPO cloning reaction was added to a vial of One Shot TOP10 compentent E. Coli cells, mixed gently, and incubated on ice for 15 minutes.  The cells were heat shocked at 42 C for 30 seconds and immediately transferred to ice.  240 uL of SOC medium was added, the eppendorf tube was capped and shaken at 37 C for 1 hour.   Next, a 1:10, 1:50 and 1:100 dilution of the cells were made in 200 uL SOC medium and spread on an LB plate with 100 ug/mL ampicillin with a sterile cell spreader.  The cells were incubated overnight at 37 C.   The next day, individual colonies were picked and cultured in 2 mL SOC medium overnight in a 5 mL loose screw capped cell culture flask.  Plasmid DNA was isolated with a plasmid DNA Miniprep kit (Invitrogen) and sent to Seqwright sequencing service for analysis.  The M13 Forward primer was used to identify the integration of DNA into the vector.  Sequences were aligned using DOSA (Daughter of Sequence Alignment) software using 80\% sequence identity, minimum n-mer of 6, and 80\%  homology.  

\vspace{.7cm}
\noindent
\textbf{
\emph{Transmission Electron Microscopy Analysis}}

Samples for TEM analysis were prepared by placing stable solutions of gold nanoparticles formed in the presence or absence of ssDNA on carbon-coated copper TEM grids. The TEM grids were allowed to stand for 2 minutes, excess solution was removed using blotting paper and the grid was allowed to dry prior to measurement. TEM measurements were performed on a Technai F20 instrument operated at an accelerating voltage at 80 kV.  Image J was used to calculate gold nanoparticle diameters of the material produced by a randomized pool, an evolved pool and without ssDNA.  The set scale tool was used to calibrate the distance in pixels that corresponded to the size of the scale bars.  Particle diameters were measured (n = 200) for each sample in order to determine the average nanoparticle diameter and standard deviation.  The data were binned in 1 nm increments in order to generate a size distribution plot of the resultant nanoparticle solutions.

\vspace{.7cm}

\noindent
\textbf{
\emph{Buffers and Cell Culture:}}

\noindent
1X PBS - 137 mM NaCl, 2.7 mM KCl, 10 mM Na$_2$HPO$_4$, 1.8 mM NaH$_2$PO$_4$ in Milli-Q H$_2$0, pH adjusted to 7.4 with 10 M NaOH.

\noindent
1X SQ20 Buffer (Primer Extension Buffer) - 120 mM Tris-HCl, pH 7.8, 100 mM KCl, 60 mM NH4SO4, 70 mM MgSO4, 1\% Triton-X100, 1 mg/mL BSA

\noindent
1X HotStart PCR Buffer-20 mM Tris-HCl (pH 8.3 at 25°C), 20 mM KCl, 5 mM (NH$_4$)$_2$SO$_4$

\noindent
SOC Medium - 10 mM NaCl, 2.5 mM KCl, 10 mM MgCl2, 2\% w/v tryptone, 0.5\% w/v yeast extract, 20 mM glucose

\bibliographystyle{ieeetr}	
\nocite{*}		
\bibliography{Bibliography}		


\end{document}